\documentclass[aps,amsmath,amssymb,reprint]{revtex4-1}
\usepackage{dcolumn}
\usepackage[utf8]{inputenc}
\usepackage[T1]{fontenc}
\usepackage{mathptmx}
\usepackage{geometry}
\usepackage[colorlinks=true]{hyperref}
\usepackage{float}
\usepackage{placeins}
\usepackage{color}
\usepackage{natbib}
\usepackage{latexsym,bm}
\usepackage{amsfonts}
\usepackage{graphicx}
\usepackage{longtable}
\begin{document}
\title{Analytical prediction of electrowetting-induced jumping motion for droplets on textured hydrophobic substrates: effects of the wetting states}

\author{Kaixuan Zhang}
\affiliation{School of Aerospace Engineering and Applied Mechanics, Tongji University, Shanghai 200092, China}

\author{Shuo Chen}
\email[Corresponding Authors:~]{schen\_tju@mail.tongji.edu.cn}
\affiliation{School of Aerospace Engineering and Applied Mechanics, Tongji University, Shanghai 200092, China}

\author{Jiayi Zhao}
\affiliation{School of Energy and Power Engineering, University of Shanghai for Science and Technology, Shanghai 200093, China}


\author{Yang Liu}
\affiliation{Department of Mechanical Engineering, The Hong Kong Polytechnic University, Hong Kong, China}

\begin{abstract}
Electric voltage applied in electrowetting can induce speading, sliding and even jumping of an individual droplet by changing the intrinsic balance of three-phase interfacial tensions, which has been widely used for droplet manipulating in microfluidics and lab-on-a-chip devices in over decades. In the present paper, we present an analytical prediction of jumping velocity for droplets electrowetting on textured hydrophobic surfaces with different wetting states. In particular, we consider a liquid droplet wetting on a textured hydrophobic substrate with a voltage applied between the droplet and the substrate. Once the voltage is turned off, the energy stored in the droplet during the electrowetting releases and could even result in the detachment of the droplet. The effects of the initial and electro- wetting states, i.e. Cassie-Baxter state and Wenzel state, on the jumping velocity of droplets are systematically discussed. Based on energy conservation, the energy conversion between the surface energy, internal viscous dissipation and the kinetic energy of droplets in different wetting states are analysed. The close-form formulas to predict the jumping velocity for different droplet wetting states are systematically derived. Finally, the unified form for predicting the electrowetting-induced jumping velocity of droplets on both flat and textured substrates with different wetting states is obtained, which can describe the jumping motion with various wetting conditions. This work provide theoretical insights on accurate control of the electrowetting-induced jumping motion of droplets on textured hydrophobic surfaces. 
\end{abstract}


\maketitle

\begin{figure}[hbpt!]
\includegraphics[width = 0.45\textwidth]{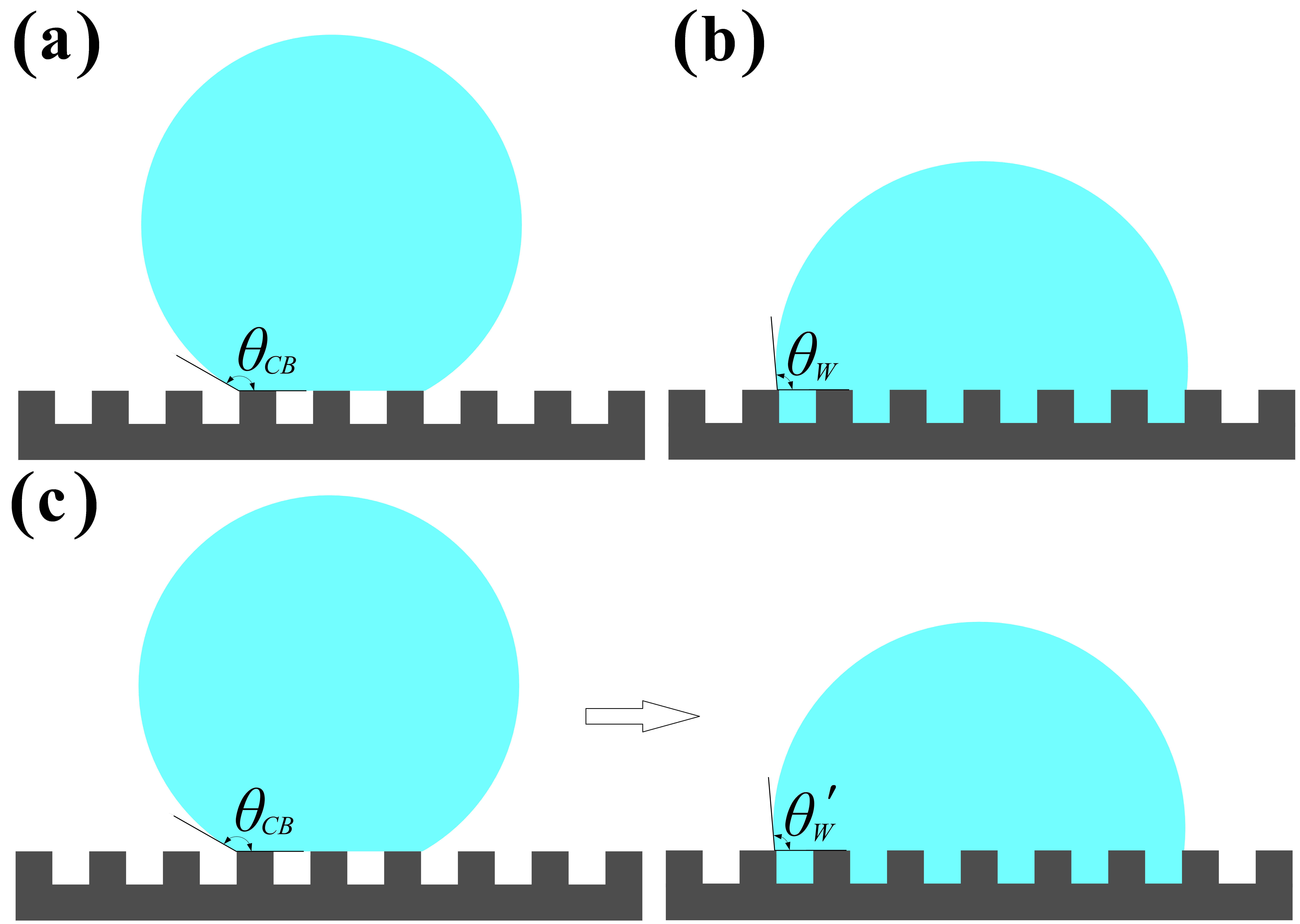}
\caption{A schematic of a droplet wetting on textured hydrophobic substrates with: (a) Cassie-Baxter state ($\theta_{CB}$), (b) Wenzel state ($\theta_{W}$)~and~(c) Wenzel state transferred from Cassie-Baxter state ($\theta_{W}^{\prime}$)}
\label{fig_cassieandwenzel}
\end{figure}

Droplet jumping on hydrophobic surfaces has attracted researchers' attentions due to its potential applications in many industrial fields, such as anti-icing~\cite{2013_Boreyko_Delayed}, anti-dew~\cite{2009_Boreyko_Self}, cleaning~\cite{2013_Wisdom_Self, 2014_Liu_self-propelled, 2014_Liu_Numerical, 2015_Enright_How, 2019_kai} and heat transfer enhancement~\cite{2013_Miljkovic_Jumping,2017_Wiedenheft_hot, 2017_Oh_Jumping}. Electrowetting is one of the most efficient techniques to manipulate the droplets to jump from hydrophobic surfaces, which has been used for accurately droplet controlling in many microfluidic applications over the past decades~\cite{2006_Bahadur_An, 2012_Li_Dissipative, 2013_Zhao_Fundamentals,2014_Arscott_Electrowetting, 2015_zhao_statics, 2017_wang_wetting, 2017_Lu_Dynamics, 2018_zhao, 2019-Johansson, raman2020electrically}. In particular, many researchers concentrate on understanding the dynamic mechanisms of electrowetting-induced jumping motion of droplets on hydrophobic substrates from analytical methods\cite{2019_Kai_pof, 2016_Cavalli_Electrically}, experimental measuraments~\cite{2016_Cavalli_Electrically, 2012_Lee_Droplet, 2014_Lee_Electrowetting, 2015_Hong_Detaching, 2016_Yan_Droplet,  2017_Wang_jumping} and numerical simulations~\cite{2016_Raman_A, 2018_Islam_a}. For instance, based on a series of experiments, Lee et al.~\cite{2012_Lee_Droplet, 2014_Lee_Electrowetting} investigated the electrowetting-induced jumping of droplets on hydrophobic surfaces, and suggested that the detachment of droplets can be improved by tuning the wettability of the substrates or enhancing the frequency of the square pulse signals. Raman et al.~\cite{2016_Raman_A} simulated the dynamic process of electrowetting-induced jumping motion based on the Lattice Boltzmann method, in which the viscous dissipation during the detachment can be increased by using higher voltage. Cavalli et al.~\cite{2016_Cavalli_Electrically} studied the electrowetting jumping of droplets from both experimental and numerical methods. They investigated the efficiency of the energy conversion between the surface energy and the gravitational potential energy of the droplet after the jumping motion occurs. Their results indicated that the finite wettability of the substrate can affect the detachment dynamics and they proposed a novel rationale for the previously reported large critical radius for droplet detachment from micro-textured substrates. Vo et al.~\cite{2019_Vo_Tran} investigated the critical conditions for jumping droplets on hydrophobic substrates in experiments and their analysis demonstrate the effects of contact-line pinning on the dynamic process of droplet electrowetting-induced jumping motion. Zhang et al.~\cite{2019_Kai_pof} derived a close-form formula to describe the energy transition during the process of droplet jumping from flat hydrophobic substrates and based on energy conversion of the droplet-substrate system, the model can accurately predict the electrowetting-induced jumping velocity of droplets on flat hydrophobic substrates with a range of wettabilities. Their prediction for droplet jumping from flat hydrophobic substrates got good agreement with the previous experimental and numerical study. They also confirmed the prediction of their theory by using many-body dissipative particle dynamics, which has been widely used to investigate microfluidic dynamic process in many applications with free interfaces~\cite{2011_Arienti, 2013_Li, 2015_Wang_Chen, 2017_jiayiZhao, 2018_Lin, 2018_Pan, 2020_Zhao_Chen}. However, A further understanding of the kinetic process on the droplets wetting on textured hydrophobic substates with different natural and even electrical wetting states is still lacking. The effects of wetting states on the accurate prediction on the escape of velocity of the droplet need a systermatically study. In this work, we present a theoretical expression of the electrowetting-induced jumping velocity of a liquid droplet on textured hydrophobic surfaces with different wetting states, which can be taken as an extension of the previous model for droplet detachment from flat hydrophobic substrates. More specifically, we consider a liquid droplet wetting on a textured hydrophobic substrate. The natural wetting state could be Cassie-Baxter state or Wenzel state, as shown in Fig.~\ref{fig_cassieandwenzel} (a) and (b). The static contact angles are marked as $\theta_{CB}$ and $\theta_W$, respectively. Fig.~\ref{fig_cassieandwenzel} (c) shows that the droplet is in Wenzel state which is tranferred from the natural Cassie-Baxter state. In that case, the value of the contact angle ($\theta_{W}^{\prime}$) is usually different from that of the natural Wenzel state. With a voltage applied between the droplet and the substrate, the Maxwell stress concentrated on the triple-phase contact line can break the intrinsic balance between three-phase interfacial tensions and deforms the droplet. By turning off the applied voltage, the energy stored in the droplet surface during the deformation can make the droplet retract and even jump from the substrate. Here, the effect of the gravity force is neglected in the derivations as the size of the droplets in our assumption is much smaller than the capillary length $l_c$, which is given by $l_c = (\gamma / \rho g)^{(1/2)}$ with $\gamma$, $g$ and $\rho$ being the liquid-vapor surface tension, the gravity and the density of the liquid. For water, $l_c = 2.7 mm$.  
Corresponding to different wetting states, the apparent contact angles can be described by the derived models from Young's equation~\cite{1805_Young}, i.e. Cassie-Baxter equation~\cite{1944Cassie} and Wenzel equation~\cite{1936Wenzel}, respectively. The surface energy of droplets is related with the three-phase interaction at the interfaces~\cite{2003_Marmur_Wetting}. 

In particular, for a droplet in Cassie-Baxter wetting state with an apparent contact angle $\theta_{CB}$, we assume the volume of the droplet is a constant value of $V_l$, the wetting radius should be $R_{C B}=\left[3 V_{l} /\left(\pi\left(2-3 \cos \theta_{C B}+\cos ^{3} \theta_{C B}\right)\right)\right]^{1 / 3}$. Then, the surface energy of the droplet in Cassie-Baxter state ${E_{CB}}$ can be described as 
\begin{equation}
\begin{aligned} E_{C B} &=\gamma A_{l v}+\gamma_{s l} A_{s l}+\gamma_{s v}\left(\Lambda-A_{s l}\right) \\ &=\gamma \pi R_{C B}^{2}\left[2\left(1-\cos \theta_{C B}\right)-\cos \theta_{C B} \sin ^{2} \theta_{C B}\right]+\gamma_{s v} \Lambda \end{aligned}
\end{equation}
where $\gamma$ is the liquid-vapor interface tension. $\gamma_{s l}$ and $\gamma_{s v}$ represent the solid-liquid and solid-vapor interface tensions, respectively. $\Lambda$ represents the total area of solid surface including the solid-vapor and solid-liquid interfaces. $A_{l v}$ and $A_{s l}$ are surface areas of liquid-vapor and solid-liquid interfaces, respectively.
For the droplet in Wenzel state with an apparent contact angle $\theta_{W}$, the wetting radius should be $R_{W}=\left[3 V_{l} /\left(\pi\left(2-3 \cos \theta_{W}+\cos ^{3} \theta_{W}\right)\right)\right]^{1 / 3}$. In this case, the surface energy $E_W$ can be described as
\begin{equation}
\begin{aligned}
E_{W} &=\gamma A_{l v}+\gamma_{s l} A_{s l}+\gamma_{s v}\left(\Lambda-A_{s l}\right) \\
&=\gamma \pi R_{W}^{2}\left[2\left(1-\cos \theta_{W}\right)-\cos \theta_{W} \sin ^{2} \theta_{W}\right]+\gamma_{s v} \Lambda
\end{aligned}
\end{equation}
As we all know, the surface energy of droplets in the Cassie-Baxter state is much higher than that in Wenzel state. If the wetting state is transferred into Wenzel state due to some external small variations, the wetting transition from Cassie-Baxter state into Wenzel state could occur. In that case, the static contact angle can become different than before, which is marked as $\theta^{\prime}$. And then the corresponding surface energy $E_{W}^{\prime}$ can be described as
\begin{equation}
\begin{aligned}
E_{W}^{\prime} &=\gamma A_{l v}+\gamma_{s l} A_{s l}+\gamma_{s v}\left(\Lambda-A_{s l}\right) \\
&=\gamma \pi {R_{W}^{\prime}}^{2}\left[2\left(1-\cos \theta_{W}^{\prime}\right)-\cos \theta_{W}^{\prime} \sin ^{2} \theta_{W}^{\prime} \right]+\gamma_{s v} \Lambda
\end{aligned}
\end{equation}  
where $R_{W}^{\prime}=\left[3 V_{l} /\left(\pi\left(2-3 \cos \theta_{W}^{\prime}+\cos ^{3} \theta_{W}^{\prime}\right)\right)\right]^{1 / 3}$ is the radius of the droplet with a spherical cap.
During electrowetting, by applying an external voltage $U$ between the droplet and the substrate, the droplet can spread on the surface and the apparent contact angle experiences a significant reduction. This is because the solid-liquid interface tension is reduced by the Maxwell stress concentrated on the triple-phase contact line\cite{2005_Quinn, 2010_Kang_Analysis}. The relationship between the spreading equilibrium state with a smaller contact angle $\theta_E$ and the heterogeneity of the textured substrates was investigated by Wang et al.~\cite{2017_wang_wetting}, which can be described by the modified Lippmann-Young equation until saturation of the contact angle occurs,
\begin{equation}
\cos \theta_{E}=f_{1}\left(\cos \theta_{Y}+\frac{1}{2} \frac{\varepsilon U^{2}}{d \gamma_{l v}}\right)-f_{2}\end{equation}
where $\varepsilon$ is the electrical permittivity and d is the thickness of the insulating layer. $f_{1}=\alpha \beta(1+\lambda)$ and $f_{2}=1-\alpha \beta$ represent the heterogeneous coefficients of the substrates. Here the substrates in our model are all isotropic, i.e. $\lambda = 1$. $\alpha$ and $\beta$ are the coefficients which represent the wetting state of droplets on textured substrates. For $\beta = 1$, the droplet is in Cassie-Baxter state, and the electrowetting contact angle is marked as $\theta_{E-CB}$. The electrowetting radius can be described as $R_{E-C B}= \left[3 V_{l} /\left(\pi\left(2-3 \cos \theta_{E-C B}+\cos ^{3} \theta_{E-C B}\right)\right)\right]^{1 / 3}$. While for $\alpha = 1$, the droplet is in Wenzel state. The electrowetting contact angle is marked as $\theta_{E-W}$. And the electrowetting radius is described as $R_{E-W}=\left[3 V_{l} /\left(\pi\left(2-3 \cos \theta_{E-W}+\cos ^{3} \theta_{E-W}\right)\right)\right]^{1 / 3}$.

When the applied voltage is turned off suddenly, since the relaxation time of the droplet is much smaller that the characteristic time of discharge process of interfacial charges, the shape of the droplet remains as the same at the apparent electrowetting contact angle $\theta_E$ without the Maxwell stress interacting on the triple-phase contact line. However, for droplets in Cassie-Baxter state initially, the wetting state after the voltage is off could be also in two types, i.e. Cassie-Baxter state or Wenzel state. While for droplets in Wenzel state initially, the wetting state should be still the same after the voltage is off. Thus, the surface energy of the droplet after the voltage $E_{EW}$ is off could be in three types:

(a)~For a droplet in Cassie-Baxter state at initial and also electrowetting stages, the surface energy $E_{CB-CB}$ can be described as
\begin{equation}\begin{aligned}
E_{CB-CB} &=\gamma A_{l v}^{\prime}+\gamma_{s l} A_{s l}^{\prime}+\gamma_{s v}\left(\Lambda-A_{s l}^{\prime}\right) \\
&=\gamma \pi R_{E-C B}^{2}\left[2\left(1-\cos \theta_{E-C B}\right)-\cos \theta_{C B} \sin ^{2} \theta_{E-C B}\right]+\gamma_{s v} \Lambda
\end{aligned}\end{equation}
where $A_{l v}^{\prime}$ and $A_{s l}^{\prime}$ represent the liquid-vapor and solid-liquid areas after the process of electrowetting. And the electrowetting contact angle is marked as $\theta_{E-C B}$. $\theta_{C B}$ is the initial contact angle, which represents the interaction between solid and liquid interface after the voltage is off. The derivation about this point has been illustrated in our previous work~\cite{2019_kai}.

(b)~For a droplet in Cassie-Baxter state initially and then transferring into Wenzel state induced by the electrowetting interaction, the surface energy of droplets $E_{CB-W}$ should be described as 
\begin{equation}\begin{aligned}
E_{CB-W} &=\gamma A_{l v}^{\prime}+\gamma_{s l} A_{s l}^{\prime}+\gamma_{s v}\left(\Lambda-A_{s l}^{\prime}\right) \\
&=\gamma \pi {R_{E-W}^{\prime}}^{2}\left[2\left(1-\cos \theta_{E-W}^{\prime} \right)-\cos \theta_{W}^{\prime} \sin ^{2} \theta_{E-W}^{\prime} \right]+\gamma_{s v} \Lambda
\end{aligned}\end{equation}
where $\theta_{W}^{\prime}$ is the contact angle in Wenzel state which is transferred from Cassie-Baxter state. $\theta_{E-W}^{\prime}$ is the electrowetting contact angle and the corresponding electrowetting radius of the droplet is $R_{E-W}^{\prime} = \left[3 V_{l} /\left(\pi\left(2-3 \cos \theta_{E-W}^{\prime}+\cos ^{3} \theta_{E-W}^{\prime}\right)\right)\right]^{1 / 3}$.

(c)~For a droplet in Wenzel state, the surface energy $E_{W-W}$ can be described as
\begin{equation}\begin{aligned}
E_{W-W} &=\gamma A_{l v}^{\prime}+\gamma_{s l} A_{s l}^{\prime}+\gamma_{s v}\left(\Lambda-A_{s l}^{\prime}\right) \\
&=\gamma \pi R_{E-W}^{2}\left[2\left(1-\cos \theta_{E-W}\right)-\cos \theta_{W} \sin ^{2} \theta_{E-W}\right]+\gamma_{s v} \Lambda
\end{aligned}\end{equation}

During the retract process of the droplet, the viscous dissipation $E_{vis}$ can be approximately estimated as~\cite{2019_Kai_pof, 2011_Wang_Size}
\begin{equation}E_{vis}=16 \pi \mu \sqrt{\frac{\gamma R_{0}^{3}}{\rho}}\end{equation}
where $R_0$ is the initial radius of the droplet. $\mu$ and $\rho$ are the viscosity and the density of the liquid, respectively. The detailed derivation and parameter analysis have been done in our previous work~\cite{2019_Kai_pof}, in which we confirmed that the viscous dissipation changes with the liquid-vapor interface tension and the initial radius of the droplet but is independent of the jumping velocity and the initial wetting contact angle. It is also noting that the interaction of the solid substrate on the triple-phase contact line and the oscillation after the detachment occurs are neglected in that assumption, which could slightly affect the prediction results for droplets wetting on textured substrates.

At the end of the retraction, if the residual kinetic energy is still larger than zero, the droplet would jump from the substrate and then, the surface energy of the droplet is estimated as 
\begin{equation}E_{\text {free}}=\gamma 4 \pi R_{0}^{2}+\gamma_{s v} \Lambda\end{equation}

From energy conservation, $E_{EW}=E_{vis}+E_{free}+E_{k}$. Corresponding to different types of initial wetting and electrowetting states, the kinetic energy of the droplet $E_{k}$ can be derived as:

(a)~For a droplet in Cassie-Baxter state at initial and also electrowetting stages,
\begin{equation}\begin{aligned}
E_{k-CB-CB}=& \gamma \pi R_{E-C B}^{2}\left[2\left(1-\cos \theta_{E-C B}\right)-\cos \theta_{C B} \sin ^{2} \theta_{E-C B}\right] \\
&-16 \pi \mu \sqrt{\frac{\gamma R_{0}^{3}}{\rho}}-\gamma 4 \pi R_{0}^{2}
\end{aligned}\end{equation}

(b)~For a droplet in Cassie-Baxter state initially and then transferring into Wenzel state induced by the electrowetting interaction,
\begin{equation}\begin{aligned}
E_{k-CB-W}=& \gamma \pi {R_{E-W}^{\prime}}^{2}\left[2\left(1-\cos \theta_{E-W}^{\prime}\right)-\cos \theta_{W}^{\prime} \sin ^{2} \theta_{E-W}^{\prime}\right] \\
&-16 \pi \mu \sqrt{\frac{\gamma R_{0}^{3}}{\rho}}-\gamma 4 \pi R_{0}^{2}
\end{aligned}\end{equation} 

(c)~For a droplet in Wenzel state,
\begin{equation}\begin{aligned}
E_{k-W-W}=& \gamma \pi R_{E-W}^{2}\left[2\left(1-\cos \theta_{E-W}\right)-\cos \theta_{W} \sin ^{2} \theta_{E-W}\right] \\
&-16 \pi \mu \sqrt{\frac{\gamma R_{0}^{3}}{\rho}}-\gamma 4 \pi R_{0}^{2}
\end{aligned}\end{equation}
Then, the jumping velocity $V_J$ of droplets for different electrowetting processes can be obtained as

(a)~For a droplet in Cassie-Baxter state at initial and also electrowetting stages,
\begin{equation}V_{CB-CB}=u\left[\frac{3}{2} A\left(\theta_{C B}, \theta_{E-C B}\right)-6(1+4 O h)\right]^{1 / 2}\end{equation}
where $A\left( {{\theta_{CB}},{\theta_{E-CB}}} \right) = \sqrt[3]{({\frac{4}{{2 - 3\cos {\theta_{E-CB}} + {{\cos }^3}{\theta_{E-CB}}}}})^2} \times \left[ {2\left( {1 - \cos {\theta_{E-CB}}} \right) - \cos {\theta_{CB}}{{\sin }^2}{\theta_{E-CB}}} \right]$ is the electrowetting coefficient.

(b)~For a droplet in Cassie-Baxter state initially and then transferring into Wenzel state induced by the electrowetting interaction,
\begin{equation}
V_{CB-W}=u\left[\frac{3}{2} A\left(\theta_{W}^{\prime}, \theta_{E-W}^{\prime}\right)-6(1+4 O h)\right]^{1 / 2}
\end{equation}
where $A\left( {{\theta_{W}^{\prime}},{\theta_{E-W}^{\prime}}} \right) = \sqrt[3]{({\frac{4}{{2 - 3\cos {\theta_{E-W}^{\prime}} + {{\cos }^3}{\theta_{E-W}^{\prime}}}}})^2} \times \left[ {2\left( {1 - \cos {\theta_{E-W}^{\prime}}} \right) - \cos {\theta_{W}^{\prime}}{{\sin }^2}{\theta_{E-W}^{\prime}}} \right]$.

(c)~For a droplet in Wenzel state,
\begin{equation}
V_{W-W}=u\left[\frac{3}{2} A\left(\theta_{W}, \theta_{E-W}\right)-6(1+4 O h)\right]^{1 / 2}
\end{equation}
where $A\left( {{\theta_{W}},{\theta_{E-W}}} \right) = \sqrt[3]{({\frac{4}{{2 - 3\cos {\theta_{E-W}} + {{\cos }^3}{\theta_{E-W}}}}})^2} \times \left[ {2\left( {1 - \cos {\theta_{E-W}}} \right) - \cos {\theta_{W}}{{\sin }^2}{\theta_{E-W}}} \right]$.

Generally, we put these formulas together and the unified form is described as,
\begin{equation}
V_{J}=u\left[\frac{3}{2} A\left(\theta, \theta_{E}\right)-6(1+4 O h)\right]^{1 / 2}
\end{equation}
where $A\left( {{\theta},{\theta_{E}}} \right) = \sqrt[3]{({\frac{4}{{2 - 3\cos {\theta_{E}} + {{\cos }^3}{\theta_{E}}}}})^2} \times \left[ {2\left( {1 - \cos {\theta_{E}}} \right) - \cos {\theta}{{\sin }^2}{\theta_{E}}} \right]$. For different wetting conditions with a natural wetting contact angle $\theta$ and an electrowetting contact angle $\theta_E$, one can obtain the corresponding electrowetting-induced jumping velocity $V_J$, i.e. for a droplet on:

i)~flat hydrophobic substrates, then $\theta = \theta_{Y}$ ($\theta_Y$ is the intrinsic contact angle of a substrate) and $\theta_E = \theta_{E-Y}$ ($\theta_{E-Y}$ is the electrowetting contact angle of the droplet on the flat substrate), which was proposed in the previous work~\cite{2019_Kai_pof};

ii)~textured substrates with natural and also electrowetting Cassie-Baxter state , then $\theta = \theta_{CB}$ and $\theta_E = \theta_{E-CB}$;

iii)~textured substrates with Cassie-Baxter state initially and then transferring into Wenzel state induced by the electrowetting interaction, then $\theta = \theta_{CB}$ and $\theta_E = \theta_{E-W}^{\prime}$;

iv)~textured substrates with Wenzel state, then $\theta = \theta_{W}$ and $\theta_E = \theta_{E-W}$.

In summary, this work provides an analytical investigation on the energy conversion between the surface energy, viscous dissipation and kinetic energy during the electrowetting-induced jumping motion for droplets on textured hydrophobic substrates with different wetting states. The theory for droplets on flat hydrophobic substrates is extended to describe the electrowetting-induced detachment of droplets on textured hydrophobic substrates, which elaborates the relationship between the electrowetting-induced velocity, the Cassie-Baxter or Wenzel contact angle, the modified Lippmann-Young contact angle, and the Oh number. The effects of natural and electrical wetting states on the jumping velocity of droplets are discussed and the corresponding analytical predictions of electrowetting-induced velocity are given. The unified form for predicting the electrowetting-induced jumping velocity of droplets on both flat and textured substrates with different wetting states is obtained, which can describe the jumping motion with various wetting conditions. It is noting that the oscillation of contact line during the retraction on the substrates and after the detachment of droplets are neglected in the assumption, whose effects need to be investigated in further research. The gravity is also neglected which indicates that this model can be used for micro- or nano- droplets whose size is much smaller than the capillary length. This work can provide new insights on accurate control of the electrowetting-induced jumping motion of droplets on textured hydrophobic substrates. 

This work was supported by the National Natural Science Foundations of China~(Grant No.\ 11872283) and Shanghai Science and Technology Talent Program (Nos. 20YF1432800).
\bibliographystyle{unsrt}
\bibliography{references}

\end{document}